\documentclass{article}

\usepackage{arxiv}

\usepackage[utf8]{inputenc} % allow utf-8 input
\usepackage[T1]{fontenc}    % use 8-bit T1 fonts
\usepackage{hyperref}       % hyperlinks
\usepackage{url}            % simple URL typesetting
\usepackage{booktabs}       % professional-quality tables
\usepackage{amsfonts}       % blackboard math symbols
\usepackage{nicefrac}       % compact symbols for 1/2, etc.
\usepackage{microtype}      % microtypography
\usepackage{lipsum}		% Can be removed after putting your text content
\usepackage{graphicx}
\usepackage{natbib}
\usepackage{doi}
\usepackage{listings}
\usepackage{subcaption}
\usepackage{tabularx}
\usepackage{breqn}
\usepackage{acro}           % acronyms
\usepackage{ragged2e}

\usepackage{todonotes}
\usepackage[normalem]{ulem}

\usepackage{soul}

\newskip\movedskip
\newcommand{\movespaceafter}[1]{%
    \movedskip=0pt%
    \ifhmode\ifdim\lastskip=0pt\else\movedskip=\lastskip\unskip\fi\fi
    #1\ifdim\movedskip=0pt\else\hskip\movedskip\fi
    \ignorespaces}

\definecolor{mossgreen}{HTML}{146614}

\lstdefinelanguage{json}{
    numberstyle=\small,
    rulecolor=\color{black},
    showspaces=false,
    showtabs=false,
    breaklines=true,
    breakatwhitespace=true,
    basicstyle=\fontsize{8}{8}\ttfamily,
    upquote=true,
    string=[b]"
}

\newcommand{\dash}{\kern -.07em\_\kern .07em}
\newcommand{\NL}{\hfil\null\penalty-9999}
\def\mymathhyphen{{\hbox{-}}}

\DeclareAcronym{EMBER}{short=EMBER,long=Elastic Malware Benchmark for Empowering Researchers}
\DeclareAcronym{SOREL}{short=SoReL,long=Sophos\slash ReversingLabs 20 million sample dataset}
\DeclareAcronym{MAEC}{short=MAEC,long=Malware Attribute Enumeration and Characterization}
\DeclareAcronym{PARCEL}{short=PARCEL,long=Parallel Class Expression Learner}

\title{Knowledge-Based Dataset for Training\\PE Malware Detection Models}

\author{ \href{https://orcid.org/0000-0002-8315-5301}{\includegraphics[scale=0.06]{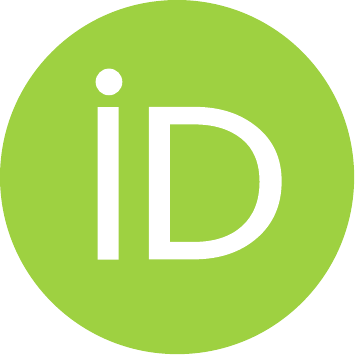}\hspace{1mm}Peter \v{S}vec} \\
	Institute of Computer Science and Mathematics\\
	Faculty of Electrical Engineering and Information Technology\\
	Slovak University of Technology\\
	Ilkovi\v{c}ova 3, Bratislava, Slovakia \\
	\texttt{peter.svec1@stuba.sk}
	\And
	\href{https://orcid.org/0000-0003-0634-9476}{\includegraphics[scale=0.06]{orcid.pdf}\hspace{1mm}\v{S}tefan Balogh} \\
	Institute of Computer Science and Mathematics\\
	Faculty of Electrical Engineering and Information Technology\\
	Slovak University of Technology\\
	 Ilkovi\v{c}ova 3, Bratislava, Slovakia \\
	\texttt{stefan.balogh@stuba.sk} \\
		\And
	\href{https://orcid.org/0000-0001-6384-9771}{\includegraphics[scale=0.06]{orcid.pdf}\hspace{1mm}Martin Homola} \\
	Department of Applied Informatics\\
	Faculty of Mathematics, Physics and Informatics\\
	Comenius University\\
	 Mlynsk{\'a} dolina, Bratislava, Slovakia \\
	\texttt{homola@fmph.uniba.sk} \\
			\And
	\href{https://orcid.org/0000-0002-3406-3574}{\includegraphics[scale=0.06]{orcid.pdf}\hspace{1mm}J{\'a}n K\v{l}uka} \\
	Department of Applied Informatics\\
	Faculty of Mathematics, Physics and Informatics\\
	Comenius University\\
	 Mlynsk{\'a} dolina, Bratislava, Slovakia \\
	\texttt{kluka@fmph.uniba.sk} \\
}

\renewcommand{\headeright}{Technical Report}
\renewcommand{\undertitle}{Technical Report}
\renewcommand{\shorttitle}{Knowledge-Based Dataset for Training PE Malware Detection Models\qquad}

\hypersetup{
pdftitle={Knowledge-Based Dataset for Training Windows Malware Detection Models},
pdfsubject={TODO,q-bio.NC, q-bio.QM},
pdfauthor={Peter \v{S}vec, \v{S}tefan Balogh, Martin Homola, J{\'a}n K\v{l}uka},
pdfkeywords={knowledge base, malware, windows, dataset},
}

\begin{document}
\maketitle

\begin{abstract}
    Ontologies are a standard for semantic schemata in many knowledge-intensive domains of human interest. They are now becoming increasingly important also in areas until very recently dominated by subsymbolic representations and machine-learning-based data processing. One such area is information security, and more specifically malware detection.
    We propose \emph{PE Malware Ontology} that offers a reusable semantic schema for Portable Executable (PE, Windows binary format) malware files. The ontology was inspired by the structure of the data in the \acs{EMBER} dataset and it currently covers the data intended for static malware analysis. With this proposal, we hope to achieve:
    a)~a unified semantic representation for PE malware datasets that are available or will be published in the future;
    (b)~applicability of symbolic, neural-symbolic, or otherwise explainable approaches in the PE Malware domain that may lead to improved interpretability of results which may now be characterized by the terms defined in the ontology; and
    (c)~by joint publishing of semantically treated \acs{EMBER} data, including fractional datasets, also improved reproducibility of experiments.
\end{abstract}

% keywords can be removed

\section{Introduction}
\label{sec:introduction}

There are currently a number of datasets that may be used to train
malware detection models. Among the most popular recent datasets we find
\acf{EMBER} \citep{anderson2018ember}
(approx.\ 1.1 million samples) and more recently \acf{SOREL}
\citep{harang2020sorel} (approx.\ 20 million samples).
The advantage of these datasets is that they were already analyzed by
security experts and for each sample they provide extensive feature data
extracted and pre-processed in structured textual format. In
addition, \ac{SOREL} provides disarmed binaries to some of the samples.
Several other online data sources are available, even some that contain
genuine malicious binaries in their samples. Such binaries may be used if
some additional information about the sample needs to be extracted.

Specifically both \ac{EMBER} and \ac{SOREL} cover the domain of PE Malware
(i.e., Windows executables and libraries).  Both are well tailored for
processing by various machine learning tools; for example \ac{EMBER}
directly provides a script to vectorize the data which is required by many
sub-symbolic classifiers.

However, in reality the data contained in these datasets is structured (JSON
files) and all textual characteristics provided for each sample can
be interpreted as meaningful properties. Such data is essentially symbolic
and can be viewed as (or, exported into) a knowledge base or a knowledge
graph. Such treatment of the data would enable to capitalize on
knowledge-processing tools, or even some tools rooted in the more recent
neural-symbolic AI movement, especially to improve the interpretability of
the resulting classifier models.

To give an example, approaches such as concept-learning may be used to obtain a symbolic characterization of a malware sample in the form of a concept expression based on an ontology \citep{svec2021experimental}.
Knowledge-base embedding \citep{TRANSE} may be used to improve effectiveness of learned models by injecting prior symbolic knowledge. More recently, neural network activation patterns recognition and alignment with ontology concepts \cite{deSousaRibeiro2021} may be used to explain and diagnose trained neural network classifiers.

Also, \ac{EMBER} and \ac{SOREL} are vast, they contain millions of samples.
Some of the studies that have been published \citep{vinayakumar2019,liu2020,ghouti2020} had to resort to reducing the
dataset of interest to a smaller size, especially if the trained
classifier is computationally demanding. However no unified methodology for
reducing the datasets has been established.

To alleviate the outlined issues, our work targets the
following goals:
\begin{description}
\item[Unified semantic representation:] To provide a unified
vocabulary for relevant PE malware characteristics that can be then used
to represent any relevant data sample regardless of its source.
\item[Interpretability of results:] To ensure that the vocabulary
provides a suitable nomenclature that is meaningful and recognized by
human users. Consequently prototype samples, concept descriptions, rules, or any
other characterizations expressed in the vocabulary would be understandable and explainable.
\item[Reproducibility of experiments:] Ensure that experiments may be
executed on datasets of different suitable sizes without the need to reduce
the full original dataset to smaller ones again and again in each study.
\end{description}

To address the first two goals, we have developed a unified ontology that
can provide a reusable semantic schema for PE malware files.  To the best
of our knowledge, such ontology was missing so far.
We have also aligned the nomenclature used to describe actions possibly performed by PE files with the
\acf{MAEC} standard \citep{maec} which is widely accepted for malware descriptions.

Our goal was to improve interpretability of the captured data. 
The ontology is partly based on the \ac{EMBER} dataset structure 
derived from the \ac{EMBER} data features, but it is not a direct
ontological mapping of this or any other individual dataset.
The emphasis is placed on data
features that are meaningful in malware characterization.
For this reason we included only those features that make sense from an expert's point of view, while other less meaningful features were left out. Some of the left-out features (e.g., file size) may even cause the trained classifiers to be prone to trial counterattacks.
Also some of the features captured in the ontology can be considered as \emph{derived} in that they combine multiple low-level EMBER data entries into a meaningful feature. Two features are based on thresholds on top of
numeric features (e.g., ``high entropy''). This is useful for systems that have difficulties with handling numeric values.

But our goal was also to create a dataset that could be effectively used as a standard benchmark to help advance learning algorithms in the malware detection domain and that would also offer unambiguous reproducibility of experimental results. To this end, we release \ac{EMBER} data (2018 version) which has been translated into RDF semantic format, including smaller fractional datasets which have been fixed for sake of better comparison of diverse experiments that require smaller data volume than the full \ac{EMBER}.

Our ontology, including the fractional datasets, and mapping scripts is available in our GitHub repository\footnote{https://github.com/orbis-security/pe-malware-ontology}.

\section{Data sources}
\label{sec:dataset}

In this section we describe in more detail the \ac{EMBER} dataset, which we used as the foundation for our ontology, along with the modifications we proposed. The main difference between the \ac{EMBER} and \ac{SOREL} datasets is that \ac{SOREL} contains more samples together with disarmed malware binaries. However, the static properties of samples that these datasets offer are almost identical (a few are missing in SoReL). Among other things, we decided on the \ac{EMBER} dataset because it is generally more used in research (261 vs.\ 32 citations).

\subsection{\Ac{EMBER}}

The dataset contains a total of 1.1 million samples and includes 400,000 malicious samples, 400,000 benign samples and 300,000 unlabeled samples, so the dataset can also be applied to unsupervised learning algorithms. The labeled samples from the dataset are also divided into training and testing sets (600,000 for training and 200,000 for testing). The dataset itself is composed of a collection of JSON objects, where each object represents data statically extracted from the PE file of one sample. A simplified example of one sample can be seen in Listing~\ref{lst:ember}. The static properties themselves are organized as follows:

\begin{description}
	\item[General file information:] This set of features is dedicated to general information about the file, such as file size, information whether the file contains a digital signature, presence of debugging symbols, presence of a TLS section, or the number of imported and exported functions.
	\item[Header information:] These are properties found in file headers such as the target architecture for which the file was compiled, linker version, various timestamps, etc.
	\item[Section information:] This set is dedicated to individual sections in the binary file. For each section dataset contains its name, content type (code, initialized or uninitialized data), various properties (such as read, write, and execute rights), or the value of the entropy of the section's content.
	\item[Imported functions:] List of imported functions organized by the DLL (here it is necessary to note that if the file imports a certain function, it does not necessarily have to be called in the code).
	\item[Exported functions:] List of exported functions. These are mostly included only in cases where the PE file is a library.
\end{description}

In addition to the categories mentioned above, the \ac{EMBER} dataset also contains other numerical properties such as byte histogram, byte-entropy histogram, or simple statistics for the strings found in the file.

\begin{lstlisting}[language=json, caption={Static features for single binary sample.}, label={lst:ember}]
{
  "sha256": "eb87d82ad7bdc1b753bf91858d2986063ebd8aabeb8e7e91c0c78db21982a0d6", 
  "md5": "aba129a3d1ba9d307dad05617f66d8e7", 
  "appeared": "2018-01",
  "label": 1,
  "avclass": "fareit",
  "histogram": [ 96506, 8328, 5582, ... ],
  "byteentropy": [0, 4229, 269, 247, ... ],
  "strings": {
    "numstrings": 7762,
    "avlength": 181.60641587219789,
    "printabledist": [591, 51, 96, 46, ... ],
    "printables": 1409629,
    "entropy": 5.037064474164528,
    "paths": 0,
    "urls": 9, 
    "registry": 0,
    "MZ": 11
  }, 
  "general": {
    "size": 2261028, 
    "vsize": 1912832, 
    "has_debug": 0, 
    "exports": 0, 
    "imports": 17,
    "has_relocations": 1, 
    "has_resources": 1, 
    "has_signature": 0, 
    "has_tls": 1,
    "symbols": 0
  },
  "header": { 
    "coff":{
      "timestamp": 708992537,
      "machine": "1386",
      "characteristics": ["CHARA_32BIT_MACHINE", "BYTES_REVERSED_LO", "EXECUTABLE_IMAGE", ... ]
    },
    "optional": {
      "subsystem": "WINDOWS_GUI", 
      "dll_characteristics": [], 
      "magic": "PE32", 
      "major_image_version": 0, 
      "minor_image_version": 0, 
      "major_linker_version": 2, 
      "minor_linker_version": 25,
      ...
    }
  },
  "section": { 
    "entry": "CODE", 
    "sections": [
      {
        "name": "CODE",
        "size": 443392,
        "entropy": 6.532932639432919,
        "vsize": 442984,
        "props": ["CNT_CODE", "MEM_EXECUTE", "MEM_READ"]
      },
    ...
    ]
  },
  "imports": {
    "kernel32.dll": ["DeleteCriticalSection", "TlsSetValue", "Sleep", ... ],
  },
  "exports": [],
  "datadirectories": [ { "name": "EXPORT_TABLE", "virtual_address": 0 }, ... ]
}
\end{lstlisting}

\subsection{Other sources}
\label{sec::othersources}

Following the success of \ac{EMBER}, \acf{SOREL} was published by \citet{harang2020sorel} who especially aimed at scaling up the volume of samples available.
The structure of \ac{SOREL} samples is similar to \ac{EMBER} and it covers the same static features with very minor differences. These once more represented in JSON.
In addition to approximately 15M JSON structured samples that are labelled and over 4M additional ones that are unlabelled, \ac{SOREL} also contains approximately 10M real binary malware samples that have been disarmed in by modifying their header file). The ontology presented in this work may be directly applied on \ac{SOREL}.

Among the most popular sources (that provide real binaries) are {VirusTotal} \citep{virustotal}, which, however, provides paid services (along with benign files), {VirusShare} \citep{virusshare}, which currently contains approx. 50 million malicious samples or {MalShare} \citep{malshare}. Known datasets that contain dynamic properties include \textsc{Malrec} \citep{severi2018m}, which captures the overall activity of malware in the form of logging all sources of indeterminism in the system, such as system calls, peripheral devices, etc. In total, it contains approximately 66,000 samples. A smaller dynamic dataset that focuses primarily on logging API calls is {Mal-API-2019} \citep{catak2019benchmark} and contains approximately 7,000 samples. In 2022, a dataset from AVAST was released \citep{bosansky2022avast}, which, unlike previous works, combines static and dynamic features and contains approx. 50,000 malware samples (its main purpose is to classify malware into individual families). Other well-known datasets include {ClaMP} \citep{clamp}, which contains only Portable Executable (PE) file headers (approx. 5,000) or {MalImg} \citep{nataraj2011malware}, which contains malicious samples, encoded in gray scale in form of approx. 10,000 samples from different families. The {DREBIN} dataset is also worth mentioning, but unlike the previous works, it contains binary samples from the Android platform \citep{arp2014drebin}.

\section{Data preprocessing}
\label{sec:preprocessing}

\subsection{Dataset standardization}
\label{sec:standardization}
% \overview{Alignment of actions etc. with standards such as \acs{MAEC}}

Individual samples may import a large number of functions from various standard DLLs implementing system APIs. They provide useful clues about the possible actions a sample can perform when running, although not all imported functions may be called and not all of these actions are relevant from the malware-detection perspective. In order to extract only relevant information from samples' imports to the ontological representation in a way that is aligned with standard best practice, we have turned to \acf{MAEC} \citep{maec}. \acs{MAEC} is a community-developed structured language for describing information about malware, combining static and dynamic features (different kinds of behavior, interactions between processes, and so on). We map the space of imported functions to \textit{malware actions}, defined in one of \acs{MAEC}'s vocabularies \citep{maec-vocabs}. This aids not only standardization of the dataset, but also achieves an effect similar to dimensionality reduction in traditional machine learning~-- irrelevant imported functions are disregarded and multiple functions with a similar effect can be mapped to a single action. The complete mapping of API functions to actions is available in our GitHub repository.

The \acs{MAEC} vocabulary describes malware actions in general, i.e., regardless of whether they are determined by static or dynamic analysis. So, in some cases, we cannot map an imported function to a \acs{MAEC} malware action. Compare, for example, the \texttt{CreateFile} and \texttt{HttpSendRequest} API calls. We can map the first one directly to the \texttt{create-file} action. However, a call to the \texttt{HttpSendRequest} function could be mapped to any of the \texttt{send-http-\textit{method}-request} actions in the vocabulary (\texttt{send-http-get-request}, \texttt{send-http-put-request}, etc.) depending on its input parameters, which are not available in the dataset.

In order to map more API calls, we have extended the MAEC actions vocabulary with additional actions summarized in Table~\ref{tab:extendedactions}. The unmappable method-specific HTTP request actions have been removed in favor of more general \texttt{send-http-request}. New actions pertaining to calls to cryptographic API functions, commonly used by malware, have been added.

Moreover, we have categorized the actions into several classes, such as networking, file access, or system manipulation, in order to aid generalization in concept learning applications of our dataset. We describe these classes in more detail in Section~\ref{sec:actions}.

\begin{table}[htbp]
	\caption{Extensions of the MAEC actions vocabulary}
	\centering
	\begin{tabular}{ll}
		\toprule
		\textbf{Action} & \textbf{Description} \\
		\midrule
		\texttt{send-http-request}
        & The action of sending an HTTP client request to a server\\
		\texttt{encrypt}
        & The action of file encryption \\
		\texttt{decrypt}
        & The action of file decryption \\
		\texttt{generate-key}
        & The action of cryptographic key generation \\
		\bottomrule
	\end{tabular}
	\label{tab:extendedactions}
\end{table}

\subsection{File and section features}
\label{sec:computed-features}

Samples in the \ac{EMBER} dataset are characterized by a number of properties of various types (cf. Lst.~\ref{lst:ember}). We have selected the most salient of them based on malware detection expert knowledge to be mapped to the ontological representation as \emph{features} of samples (PE files) or samples' sections. There are three kinds of such features in the ontology:
\begin{enumerate}
\item direct representations of boolean (actually 0/1) properties from the original dataset;
\item features obtained by pre-processing original properties not directly represented in the ontology;
\item features obtained by pre-processing properties that are also represented directly.
\end{enumerate}

Features of the first kind arise from sample properties such as \texttt{general.has\_relocations}. The original properties mapped to these features are listed within the description of features in Sec.~\ref{sec:filefeatures}.

There are two features of the second kind: One of them is the presence of a non-empty Common Language Runtime data directory (in the \texttt{datadirectories} property) in the sample, which is indicative of .NET binaries. The other is the sample's entry point being located in a non-executable section (checked by inspecting the \texttt{section.entry} property and the respective member of the \texttt{section.sections} array). Machine-learning algorithms can thus take advantage of these two important features even though the underlying properties are not represented in the ontology directly.

Features of the third kind, also called \emph{derived} features, can be defined by OWL~2 expressions, but they are known to experts to be important malware indicators and are thus worth naming explicitly. Moreover, assigning these features to samples during dataset pre-processing exposes them even to tools that lack the required expressivity (cardinality restrictions, data type restrictions, enumerated data types with long lists of literals) or enabling it imposes a significant performance penalty. If, e.g., a classifier trained on the data set is capable of working efficiently with constructs required to define some of the derived features, these may be considered redundant and the user of our dataset may instruct the tool to selectively ignore them so as not to enlarge the dimensionality or search space unnecessarily. Likewise, if the user targets a learning algorithm to learn from the derived features, e.g., for sake of efficiency, it may be indicated to remove or to ignore the original data from which they were constructed.

Two features derived from numerical sample properties in the \ac{EMBER} dataset stand out due to their important role in malware detection and a less trivial way of determining a suitable threshold controlling the assignment of the feature to a sample. These features are \emph{a low number of imports} of the sample and \emph{a high entropy} of a sample's section. They indicate that the sample may be packed. The threshold values for these features are inspired by \textit{pestudio} \citep{pestudio}, a tool commonly used by security teams for the initial assessment of malware samples. Based on \textit{pestudio}'s defaults, a sample importing less than 10 functions is considered having a low number of imports, and a section with entropy greater than~7.0 is marked as having a high entropy.

\begin{figure}[htbp]
\begin{subfigure}{.5\textwidth}
  \centering
  \includegraphics[width=.8\linewidth]{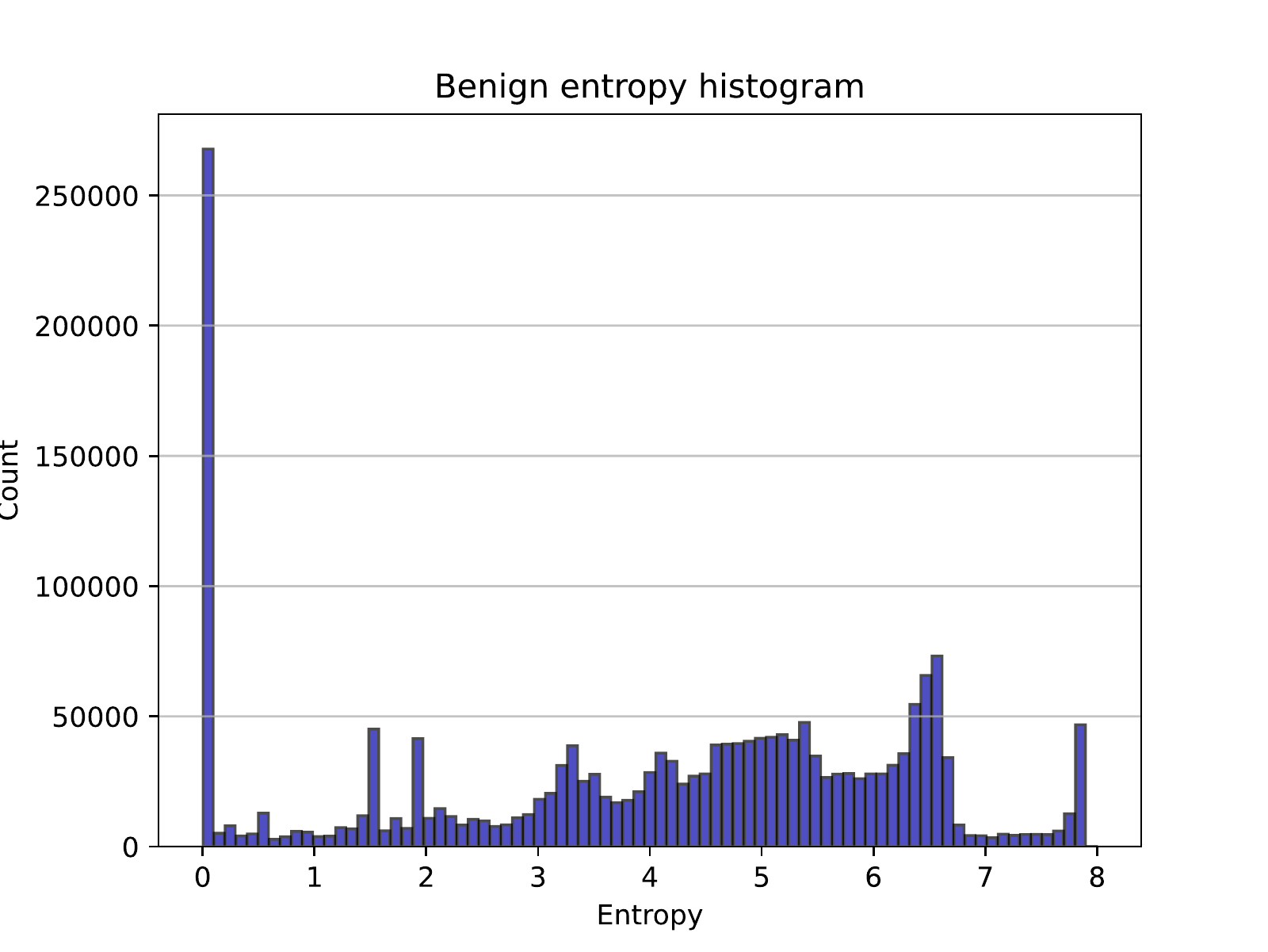}  
  \caption{Benign entropy histogram}
  \label{fig:hist-entropy-benign}
\end{subfigure}
\begin{subfigure}{.5\textwidth}
  \centering
  % include fourth image
  \includegraphics[width=.8\linewidth]{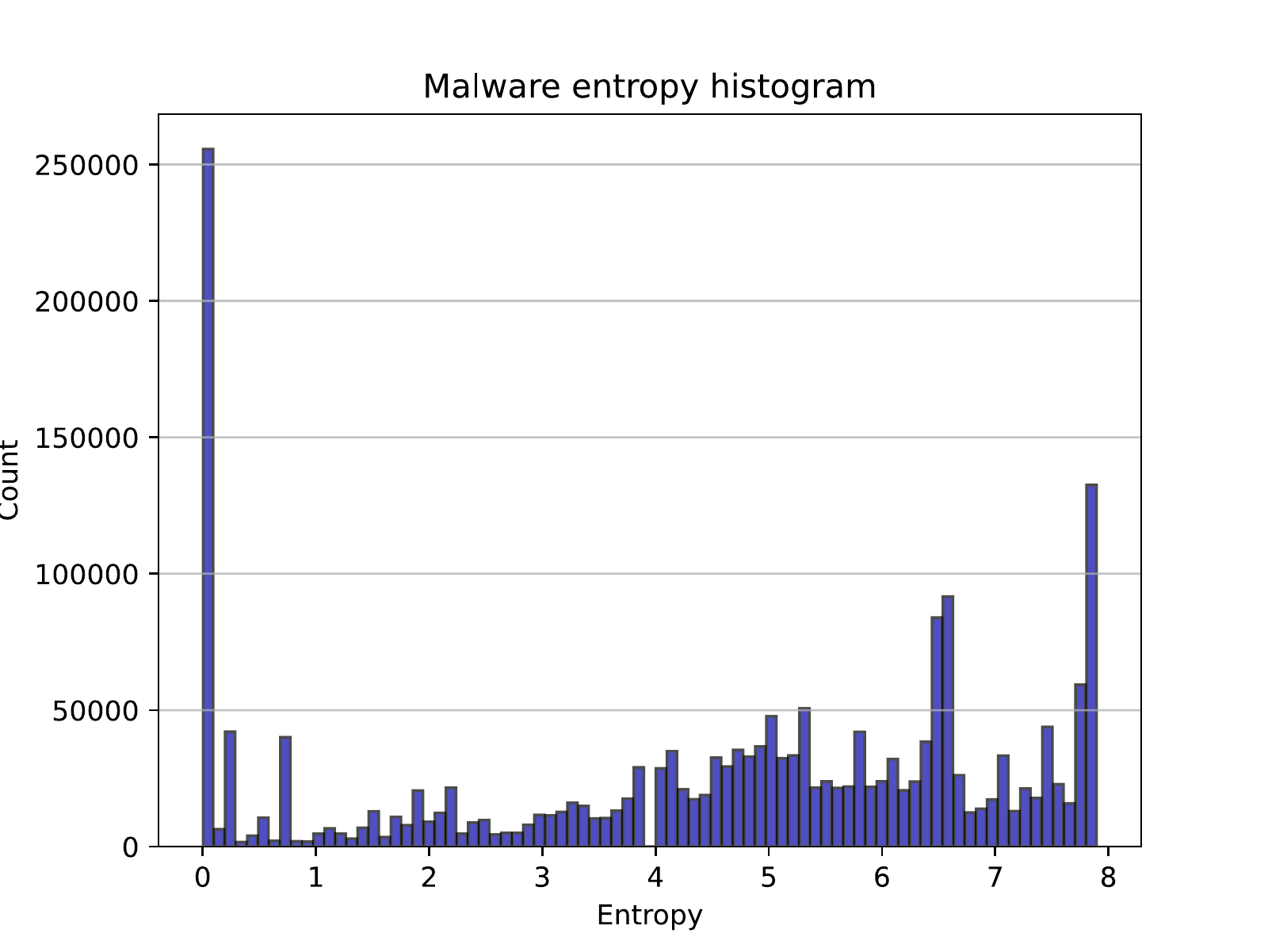}  
  \caption{Malware entropy histogram}
  \label{fig:hist-entropy-malware}
\end{subfigure}

\leavevmode%\newline

\begin{subfigure}{.5\textwidth}
  \centering
  \includegraphics[width=.8\linewidth]{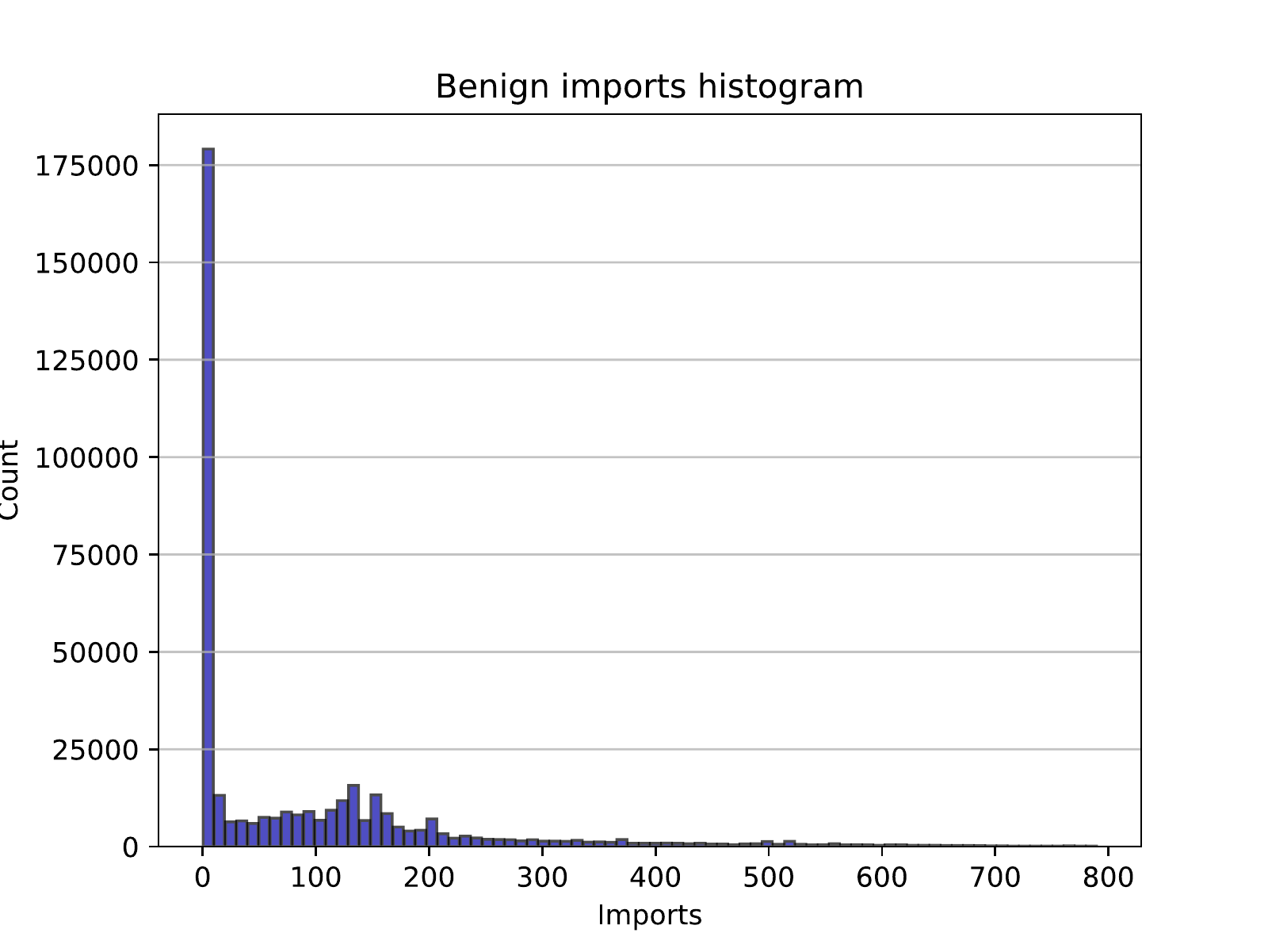}  
  \caption{Benign imports histogram}
  \label{fig:hist-imports-benign}
\end{subfigure}
\begin{subfigure}{.5\textwidth}
  \centering
  \includegraphics[width=.8\linewidth]{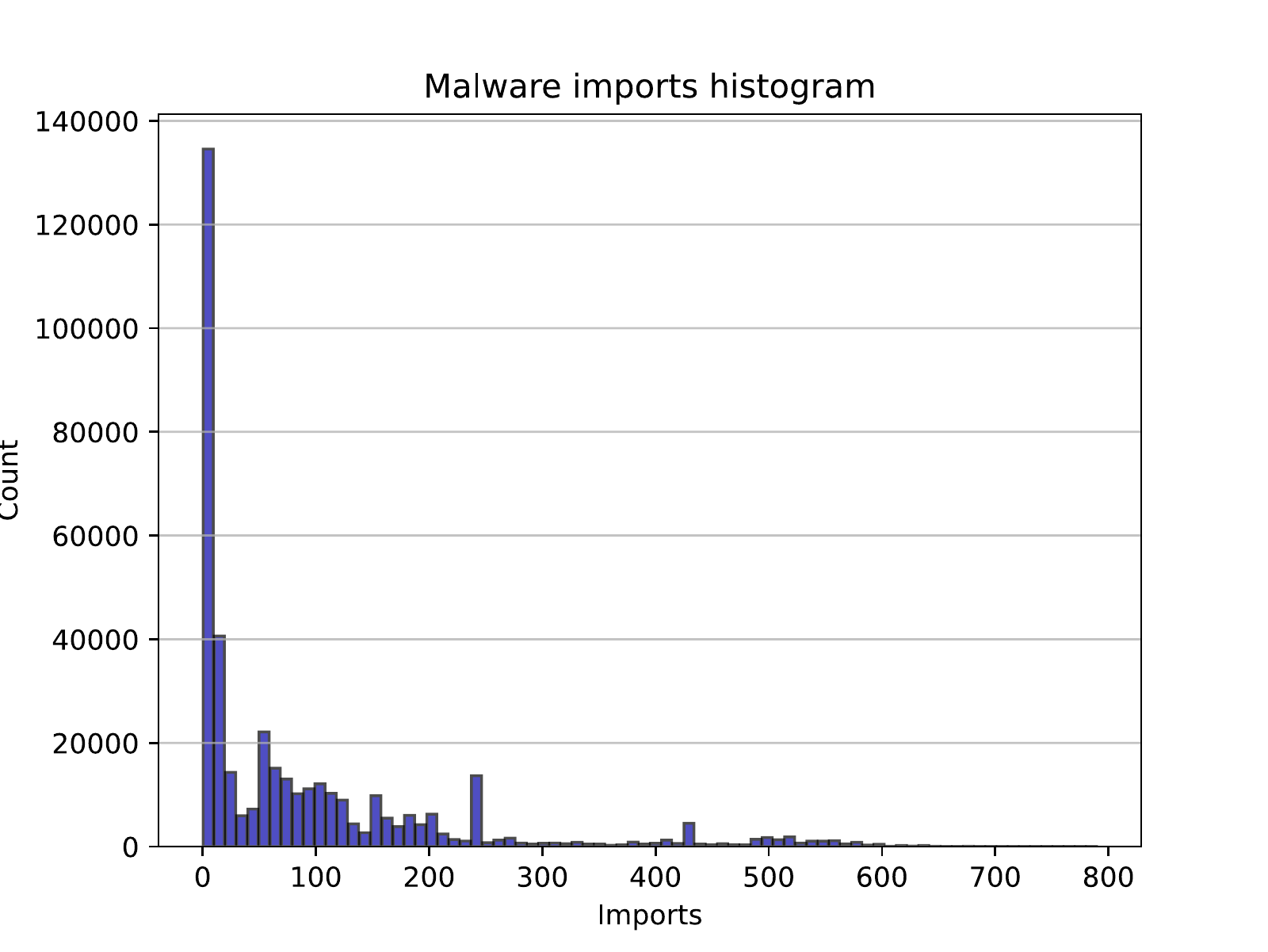}
  \caption{Malware imports histogram}
  \label{fig:hist-imports-malware}
\end{subfigure}
\caption{Entropy and imports histograms for \ac{EMBER} dataset}
\label{fig:histograms}
\end{figure}

In order to verify the threshold values from the \textit{pestudio} tool, we performed a basic statistical analysis for the entire EMBER dataset. Specifically, we were interested in histograms representing the number of imports for each sample and the entropy of each section in the dataset. The results are depicted in Fig.~\ref{fig:histograms}. The threshold value for section entropy was confirmed. We can see that there are significantly more sections in malware samples with entropy higher than~7.0 compared to the number of such sections in benign samples.
As for the number of imports, the histograms in Figs.~\ref{fig:hist-imports-benign} and~\ref{fig:hist-imports-malware} do not show any clear indicative value that separates a significant volume of malware and benign samples. In particular, a large number of both kinds of samples import less than~10 API functions. This feature alone is thus not a strong indicator of a sample's maliciousness, although it may become useful in combination with other features. Hence we decided to include it anyway and keep \emph{pestudio}'s preset threshold of~10.

Another non-trivially derived feature of sections is \emph{a non-standard name}. While we map section names to the ontological representation, describing this feature in OWL~2 requires a data property restriction with a negated list of section names usually produced by compilers and linkers. We consider it improbable that a learning algorithm would discover such a restriction. We have thus opted to collect usual section names and to add this feature to sections when EMBER data is transformed to the ontology. The list of usual section names is available in our GitHub repository.

% \subsection{Derived features}

% We have several derived features in our ontology. These are not present directly in the \ac{EMBER} dataset, but they are combination of multiple features instead. Their main purpose is to speed up learning process in concept learning algorithms. One example of derived feature is \texttt{MultipleExecutableSections} class. Information about how many sections specific sample contains and how many of them are executable is present in the ontology and it is theoretically possible for algorithms to learn them. By directly using such feature, we can speed up the learning process. Other similar features are \texttt{NonexecutableEntryPoint} or \texttt{NonstandardMZ}. \texttt{NonexecutableEntryPoint} represents if program entry point is located in a non-executable section. Information, in which section the entry point is located and what permissions it contains can be found in the \ac{EMBER} dataset. The \texttt{NonstandardMZ} feature works similarly. In the \ac{EMBER} dataset, we have information about how many MZ headers given sample contains (if it contains more than one, it may indicate an embedded PE file). The \texttt{NonstandardMZ} class therefore indicates whether a given sample contains more than one MZ header. Other derived features are discussed in Section \ref{sec:filefeatures} and Section \ref{sec:sectionfeatures} respectively.

\section{PE Malware Ontology}
\label{sec:ontology}

\begin{figure}
    \centering
    \includegraphics{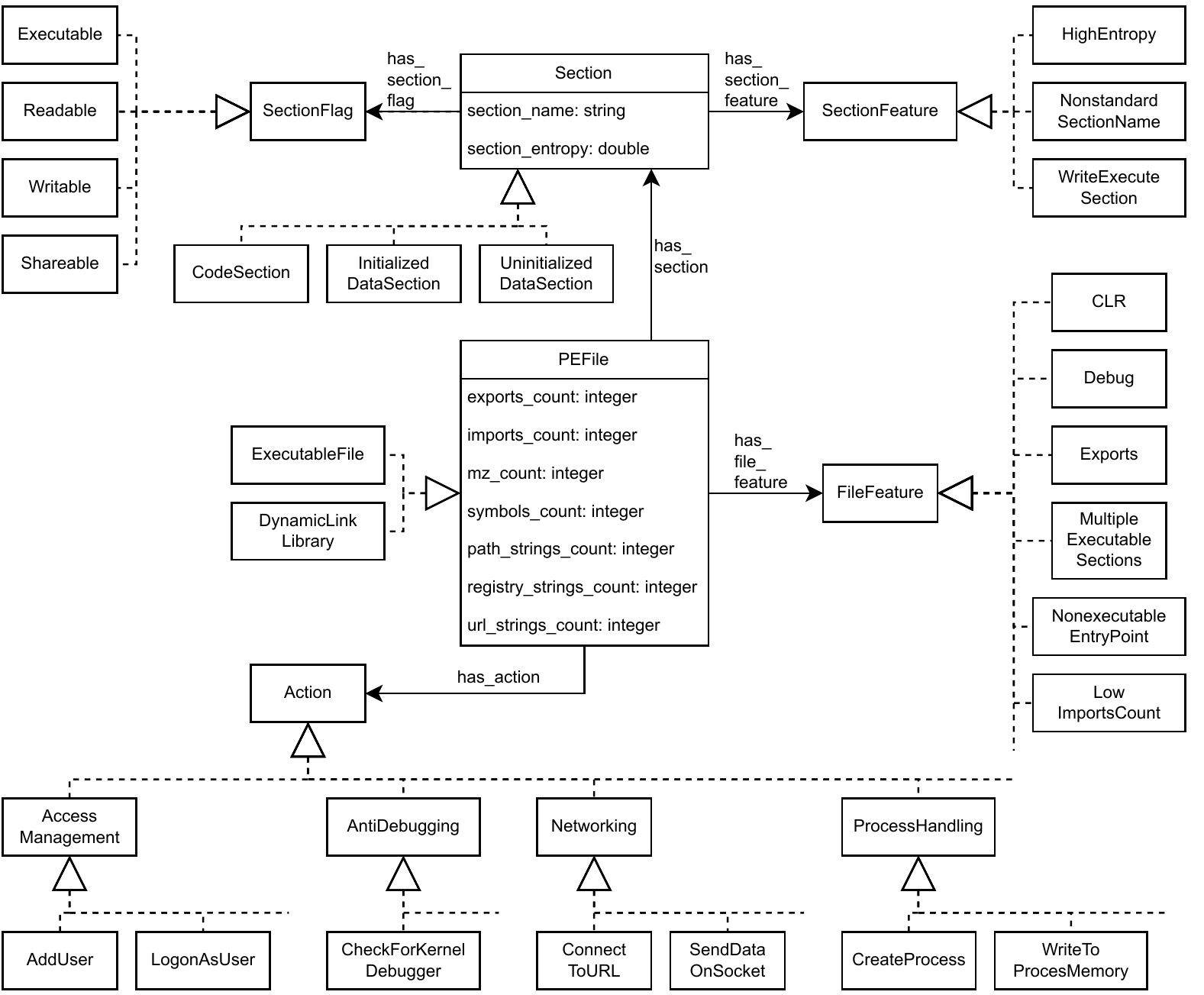}
    \caption{Core classes of the \ac{EMBER} ontology and their properties}
    \label{fig:ontology}
\end{figure}

We now introduce PE Malware Ontology. While we used the structure of the \ac{EMBER} as a starting point, it should be noted that our ontology is not a direct mapping of this or any other individual dataset.  
Since one of the most important goals was interpretability, we included only those features that make sense from an expert's point of view. For this reason, we did not use various features such as file sizes, linker versions, byte histograms, etc. In theory, there may be patterns between in features and various classifiers can learn them, but from an expert's perspective, such features are meaningless for distinguishing between malware and benign samples. Also, the use of such features could subsequently cause the learned classifiers being prone to trivial adversarial attacks. 

Our ontology is partially depicted in Figure~\ref{fig:ontology}. The core classes and properties are fully depicted, but only an illustrative sample of file features and actions is shown for brevity. The ontology contains 195~classes, 6~object properties, and 9~data properties in total.

\subsection{PE files}
\label{sec:pefiles}

Each sample of a Portable Executable file \citep{bridge2022pe} in the original dataset is represented by an instance of the \texttt{PEFile} class, central to our ontology. Each such instance is further classified either as a proper executable in the \texttt{ExecutableFile} subclass of \texttt{PEFile} or as a shared library (DLLs) in the \texttt{DynamicLinkLibrary} subclass. The respective subclass is determined based on the COFF header available in the \ac{EMBER} dataset.

The \texttt{PEFile} class is the domain of 7~data properties (see Tab.~\ref{tab:PEFile}) which provide information on the number of imports and exports, symbols, and MZ headers (mapped from the \texttt{general} property of JSON descriptions of samples), as well as the the number of path, registry, and URL strings (mapped from the \texttt{strings} property).

PE files structurally consist of sections, possess features relevant to malware detection, and may perform actions (based on their imported functions) when executed. These entities are represented by instances of the respective classes in the ontology (\verb+Section+, \verb+FileFeature+, \verb+Action+) which are linked to a \texttt{PEFile} instance via the \verb+has_section+, \verb+has_file_feature+, and \verb+has_action+ object property, respectively.

\begin{table}[htbp]
	\caption{\texttt{PEFile} properties}\label{tab:PEFile}
	\centering
    \begin{tabular}[t]{ >{\tt} l >{\tt} l @{\hspace{2\arraycolsep}} }
    \toprule
    \bf Data property & \bf Range
    \\
    \midrule
    exports\_count & xsd:integer
    \\
    imports\_count & xsd:integer
    \\
    mz\_count & xsd:integer
    \\
    path\_strings\_count & xsd:integer
    \\
    symbols\_count & xsd:integer
    \\
    registry\_strings\_count & xsd:integer
    \\
    url\_strings\_count & xsd:integer
    \\
    \bottomrule
    \end{tabular}%
    \begin{tabular}[t]{ >{\tt} l >{\tt} l @{\hspace{2\arraycolsep}} }
    \toprule
    \bf Object property & \bf Range
    \\
    \midrule
    has\_action & Action
    \\
    has\_file\_feature & FileFeature
    \\
    has\_section & Section
    \\
    \\
    \\
    \\
    \\
    \bottomrule
    \end{tabular}
\end{table}

\subsection{File features}
\label{sec:filefeatures} 

The \texttt{FileFeature} class has 15~subclasses representing various qualitative features of PE files that are, based on expert knowledge, relevant to malware detection. They are enumerated and described in Table~\ref{tab:filefeatures}. 
As detailed in Sect.~\ref{sec:computed-features}, these features can be sorted into three categories: directly represented features, pre-processed features, and additional derived features that are obtained from underlying values that are also represented.

Each subclass of \texttt{FileFeature} has a single prototypical instance. All \verb+PEFile+ instances that possess a feature are linked via the \verb+has_file_feature+ object property to the feature's prototypical instance.
Although this way of modeling features may be uncommon, it allows for easily readable descriptions of possible malware in the Manchester syntax, e.g., \texttt{ExecutableFile has_file_feature some \{multiple\_executable\_sections\}} -- to be read e.g.\,``an executable file with the feature of having multiple executable sections''.

\begin{table}[htbp]
	\caption{PE file features}
	\centering
	\begin{tabularx}{\textwidth}{l >{\RaggedRight\hangindent=1.5em} X}
		\toprule
		\textbf{File feature class}
        & \textbf{Description (sample property in JSON or equivalent OWL~2 expression)} \\
		\midrule
        \addlinespace
        \textbf{Directly represented features}
        \\\addlinespace[.5\defaultaddspace]
		\texttt{Debug}
        & Contains a debug section
            (\texttt{general.has\_debug}) \\
		\texttt{Relocations}
        & Contains a relocation section
            (\texttt{general.has\_relocations}) \\
		\texttt{Resources}
        & Contains resources (fonts, images, etc.;
            \texttt{general.has\_resources}) \\
	    \texttt{Signature}
        & Is digitally signed
            (\texttt{general.has\_signature}) \\
		\texttt{TLS}
        & Includes a Thread Location Storage section
            (possibly a secret entry point;
            \texttt{general.has\_tls})
        \\\addlinespace
		\textbf{Pre-processed features}
        \\\addlinespace[.5\defaultaddspace]
        \texttt{CLR}
        & Contains a Common Language Runtime data directory\NL
            (used in .NET binaries) \\
		\texttt{NonexecutableEntryPoint}
        & Entry point is not in an executable section
        \\\addlinespace
		\textbf{Derived features}
        \\\addlinespace[.5\defaultaddspace]
	    \texttt{Exports}
        & Exports functions (mostly in DLLs;
            \texttt{exports\_count some xsd:integer[> 0]}) \\
		\texttt{MultipleExecutableSections}
        & Has multiple sections with the executable flag
            (\texttt{has\_section min 2 (has_section\_flag some Executable)}) \\
		\texttt{LowImportsCount}
        & The number of imported functions is smaller than the threshold value\NL
            (\texttt{imports\_count some xsd:integer[< \textit{imports\_threshold}]}) \\
		\texttt{NonstandardMZ}
        & Has no or more than one MZ headers
            (possibly contains an embedded PE file;
            \texttt{mz\_count some xsd:integer[> 1]})     \\
		\texttt{PathStrings}
        & Contains strings defining paths
            (\texttt{path\_strings\_count some xsd:integer[> 0]}) \\
		\texttt{RegistryStrings}
        & Contains strings defining registry keys
            (\texttt{registry\_strings\_count some xsd:integer[> 0]}) \\
		\texttt{Symbols}
        & Has COFF debug symbols
            (deprecated in executables;
            \texttt{symbols\_count some xsd:integer[> 0]}) \\
	    \texttt{URLStrings}
        & Contains strings defining URLs
            (\texttt{url\_strings\_count some xsd:integer[> 0]}) \\
		\bottomrule
	\end{tabularx}
	\label{tab:filefeatures}
\end{table}

\subsection{Sections}
\label{sec:sections}

Instances of the \texttt{Section} class represent sections of PE file samples in the original dataset. A \texttt{PEFile} instance is linked to all its sections via the \texttt{has\_section} property. Sections are further classified based on the type of data found in the given section into one of \verb+Section+'s three subclasses: \texttt{CodeSection}, \texttt{InitializedDataSection}, and \texttt{UninitializedDataSection}.

\texttt{Section} instances have data properties assigning them their names and the entropies of their content. Furthermore, they are linked to their permission flags and features by the respective object properties shown in Table~\ref{tab:Section}.

\begin{table}[htbp]
	\caption{\texttt{Section} properties}\label{tab:Section}
	\centering
    \begin{tabular}[t]{ >{\tt} l >{\tt} l @{\hspace{2\arraycolsep}} }
    \toprule
    \bf Data property & \bf Range
    \\
    \midrule
    section\_entropy & xsd:double
    \\
    section\_name & xsd:string
    \\
    \bottomrule
    \end{tabular}%
    \begin{tabular}[t]{ @{\hspace{2\arraycolsep}} >{\tt} l >{\tt} l }
    \toprule
    \bf Object property & \bf Range
    \\
    \midrule
    has\_section\_feature & SectionFeature
    \\
    has\_section\_flag & SectionFlag
    \\
    \bottomrule
    \end{tabular}
\end{table}

\subsection{Section flags}
\label{sec:sectionflags}

The \texttt{SectionFlag} class classifies the flags given to PE file sections. In the \ac{EMBER} dataset, these flags are included in the \verb+props+ property of each section. From the malware-detection point of view, the most interesting of them are the flags controlling how processes executing the PE file may access (read, write, execute) and share the memory region into which this section is mapped. These flags can be, in theory, combined arbitrarily. Only these four flags are represented in the ontology by the \verb+Executable+, \verb+Readable+, \verb+Writable+, and \verb+Shareable+ subclasses of \texttt{SectionFlag}. Similarly to file features, each subclass has a prototypical instance that sections with the respective flag are linked to via the \verb+has_section_flag+ object property.

\subsection{Section features}
\label{sec:sectionfeatures}

Similarly to \verb+FileFeature+, \verb+SectionFeature+'s subclasses represent features of the sections that, based on our expert knowledge, are relevant for malware detection. They are listed in Table~\ref{tab:sectionfeatures}. All of these features are derived, and their values are computed from the section's data properties and flags, as detailed in Sect.~\ref{sec:computed-features}.

Instances of \texttt{Section} are linked to the prototypical instances of the three subclasses of \texttt{SectionFeature} using the \texttt{has\_section\_feature} property.

\begin{table}[htbp]
	\caption{Section features}
	\centering
	\begin{tabularx}{\textwidth}{l >{\RaggedRight\hangindent=1.5em} X}
		\toprule
		\textbf{Section feature class}
        & \textbf{Description (equivalent OWL~2 expression)} \\
		\midrule
		\texttt{HighEntropy} &
            The value of section's entropy is larger than the threshold value
            (\texttt{section\_entropy some xsd:double[> \textit{entropy\_threshold}]})   \\
		\texttt{NonstandardSectionName}
        & Section's name is not in the list of standard section names
            (\texttt{section\_name some not \{ ".text", ".data", ".rsrc", \ldots\ \}})\\
	    \texttt{WriteExecuteSection}
        & Section has write and execute permissions
            (\texttt{(has\_section\_flag some Writable) and
                (has\_section\_flag some Executable)})\\
		\bottomrule
	\end{tabularx}
	\label{tab:sectionfeatures}
\end{table}

\subsection{Annotation of derived features}
\label{sec:derivedfeaturesannot}

As discussed in Sec.~\ref{sec:computed-features}, derived features of PE files and sections may be considered redundant in some applications and retaining them in the data set may have adverse effects. We have documented such features in the ontology with the \texttt{derived\_as} annotation property, in order to facilitate their automatic identification. Each derived feature class is annotated using this property with the respective defining OWL~2 expression (see Tables \ref{tab:filefeatures} and~\ref{tab:sectionfeatures}), encoded as a string in the Manchester syntax.

\subsection{Actions}
\label{sec:actions}

Instances of class \texttt{Action} represent actions that may be taken by a process executing the code from a PE file. Actions are classified in 139~leaf subclasses of \texttt{Action} (a \emph{leaf} class is one with no named subclasses except itself) that are aligned with the \ac{MAEC} standard's Malware actions vocabulary as described in Sec.~\ref{sec:standardization}. In order to aid generalization during classifier learning, we have added 17~categories of actions listed in Table~\ref{tab:actions}. Each category is represented by an intermediate class (e.g., \texttt{Process\allowbreak Handling}; see Fig.~\ref{fig:ontology})~-- a direct subclass of \texttt{Action} and a direct superclass of the leaf \ac{MAEC}-based action classes falling into the category (e.g., \texttt{Create\allowbreak Process}, \texttt{Write\allowbreak To\allowbreak Process\allowbreak Memory}).

Currently, each leaf action class has a prototypical instance and all \verb+PEFile+ instances that may perform this action are connected to it by the \verb+has_action+ object property. This connection is created by mapping imported functions from the \verb+imports+ property of samples in the \ac{EMBER} dataset to the corresponding \ac{MAEC} malware actions if possible. For example, when generating an ontological representation of a sample importing an API function that creates a process, a prototypical instance \texttt{create-process} of the \texttt{CreateProcess} class is linked to the sample's \texttt{PEFile} instance via the \texttt{has\_action} property. Of course, importing such an API function does not necessarily mean that it will actually be called (thus creating a process) when this sample is run.

Alternatively, if dynamic tracing data from running samples in a sandboxed environment are available, every action actually performed by a sample could be represented by an individual, non-prototypical instance of \texttt{Action}, classified in the appropriate leaf subclass and further characterized by additional properties, such as parameters, previous and next action, or timestamp. Extending our ontology to cover such dynamic data may be interesting future work. However, to the best of our knowledge, no such dataset is currently available.

% Actions are created by mapping imported functions from the \ac{EMBER} dataset. The \texttt{Action} class contains 17~subclasses. The list of individual subclasses together with their description can be seen in Table \ref{tab:actions}. It is important to note that we created these classes ourselves and they are not part of the standard (we simply grouped these actions into various categories). As an example, we can show the \texttt{ProcessHandling} class. This class contains another 16 subclasses such as \texttt{CreateProcess} or \texttt{WriteToProcessMemory}. If the sample imports an API call to create a process (which does not necessarily mean that this sample will call this function and thus creating a process), then a new instance of the \texttt{CreateProcess} class called \texttt{create-process} (according to the \ac{MAEC} vocabulary) is created in our ontology and is connected to the \texttt{PEFile} class via the \texttt{has\_action} property . It should also be noted that if multiple samples import an API call for process creation, each sample will be attached to the same \texttt{create-process} instance (since it is not necessary to create a unique instance for each sample).

\begin{table}[htbp]
	\caption{Action classes}
	\centering
	\begin{tabularx}{\textwidth}{l >{\RaggedRight\hangindent=1.5em} X}
		\toprule
		\textbf{Action class}    & \textbf{Description}  \\
		\midrule
		\texttt{AccessManagement} & Managing users on the system (adding new user, enumerating existing users, etc.)    \\
		\texttt{AntiDebugging}     & Debugger detection techniques  \\
	    \texttt{Cryptography}    &  Encrypting/decrypting files, generating keys, etc.      \\
	    \texttt{DirectoryHandling}    & Manipulating with directories (creation, deletion, etc.)       \\
	    \texttt{DiskManagement}    & Mounting/unmounting disks, enumerating existing disks       \\
	    \texttt{FileHandling}    & Manipulating with files (creation, deletion, etc.)       \\
	    \texttt{InterProcessCommunication}    & Communication between processes (named pipes)       \\
	    \texttt{LibraryHandling}    & Loading library into running processes       \\
	    \texttt{Networking}    & Various networking activities (connecting to a socket, sending DNS requests)       \\
	    \texttt{ProcessHandling}    &  Various APIs for process handling, including creation or modifying process memory      \\
	    \texttt{RegistryHandling}    & Enumerating registry keys, writing new values, etc.       \\
	    \texttt{ResourceSharing}    & Manipulating with resources shared over network       \\
	    \texttt{ServiceHandling}    & Manipulating with systems's services       \\
	    \texttt{SynchronizationPrimitivesHandling}    & Handling mutexes/semaphores       \\
	    \texttt{SystemManipulation}    & Various APIs for obtaining system information       \\
		\texttt{ThreadHandling}    & Creating remote threads, enumerating running threads       \\
		\texttt{WindowHandling}    & Window manipulation APIs (creating new windows, dialog boxes, etc.)       \\
		\bottomrule
	\end{tabularx}
	\label{tab:actions}
\end{table}

\section{Fractional datasets}
\label{sec:fractions}

One of the important goals was to create datasets that would allow unambiguous reproduction of experimental results. For this reason, we generated multiple datasets with a clear definition of which samples belong to the training set and which belong to the testing set. Each dataset contains 50\% positive samples (malware) and 50\% negative samples (benign). We generated datasets of various sizes from 1000 to 800,000 (i.e. the entire \ac{EMBER} dataset excluding unlabeled samples). The main reason was that concept learning algorithms are computationally more demanding than some traditional machine learning algorithms, and for that reason it can be useful to have less robust datasets available. We also generated several variants for each fractional dataset size, which were randomly selected from the entire \ac{EMBER} dataset. More details can be seen in Table \ref{tab:fractionaldatasets}.

\begin{table}[htbp]
	\caption{Fractional datasets properties}
	\centering
%	\begin{tabular}{lrrrrrr}
%		\toprule
%		& & \multicolumn{2}{c}{\textbf{Training samples}} & \multicolumn{2}{c}{\textbf{Testing samples}} & \\
%        \cmidrule(lr){3-4}\cmidrule(lr){5-6}
%		\textbf{Name}    & \textbf{Total samples} &  \textbf{Positive} & \textbf{Negative} & \textbf{Positive} & \textbf{Negative} & \textbf{Variants} \\
%		\midrule
%		\texttt{dataset\_N\_1000.owl} & 1000    & 400 & 400 & 100 & 100 & 10\\
%		\texttt{dataset\_N\_10000.owl}     & 10,000 & 4000 & 4000 & 1000 & 1000 & 10 \\
%	    \texttt{dataset\_N\_100000.owl}   & 100,000  & 40,000 & 40,000 & 10,000 & 10,000 & 10  \\
%	    \texttt{dataset\_N\_800000.owl}   & 800,000  & 320,000 & 320,000 & 80,000 & 80,000 & 1 \\
%		\bottomrule
%	\end{tabular}
	\begin{tabular}{lrrrr}
		\toprule
		\textbf{Name}    & \textbf{Total samples} &  \textbf{Positive} & \textbf{Negative} & \textbf{Variants} \\
		\midrule
		\texttt{dataset\_N\_1000.owl} & 1000    & 500 & 500 & 10\\
		\texttt{dataset\_N\_10000.owl}     & 10,000 & 5000 & 5000 & 10 \\
	    \texttt{dataset\_N\_100000.owl}   & 100,000  & 50,000 & 50,000 & 10  \\
	    \texttt{dataset\_N\_800000.owl}   & 800,000  & 400,000 & 400,000 & 1 \\
		\bottomrule
	\end{tabular}
	\label{tab:fractionaldatasets}
\end{table}

There are three files for each individual dataset:

\begin{description}
	\item[\texttt{dataset\_N\_1000.owl}:] this is an ontology filled with individuals of OWL type, which contains 1000 malware/benign samples;	\item[\texttt{dataset\_N\_1000\_raw.json}:] original JSON samples from the \ac{EMBER} dataset (including unused features) from which the ontological dataset was generated;
    \item[\texttt{dataset\_N\_1000\_examples.json}:] list of positive and negative samples in the given dataset in JSON format.
\end{description}

We refrain from publishing a predefined split of the datsets into the training and testing part, as committing to one split may lead into lucky or unlucky setting for certain learning algorithms or even to overfitting on the given predefined split. Instead we propose to follow the $k$-fold cross-evaluation methodology \citep{kfold} in which the dataset is split in $k$ equal parts and in each run one of them is used for testing while the rest of the data is used for training. The results are then averaged. Fivefold and tenfold splits are commonly used.

\section{Usage in concept learning}
\label{sec:experiments}

The ontology can be used by various machine learning algorithms where the goal is to build a malware detection model. One such example includes the family of concept learning algorithms. Which can be used to provide descriptive characterizations of (parts of) the malware sample using therms from the ontology. Three examples of class expressions (in standard description logics syntax \citep{dlhb}) learned using \acf{PARCEL}
\cite{tran2012approach} algorithm are shown below: 

\begin{dmath}
\label{concept1}
(\exists has \dash file \dash feature.\{multiple \dash executable \dash sections\}) \sqcap (\geq 2 has \dash section.(\exists has \dash section \dash feature.\{high \dash entropy\})
\end{dmath}

\begin{dmath}
\label{concept2}
(\exists has \dash action.\{read \mymathhyphen from \mymathhyphen process \mymathhyphen memory\}) \sqcap (\exists has \dash section.(\exists has \dash section \dash feature.\{ write \dash execute \dash section\})
\end{dmath}

\begin{dmath}
\label{concept3}
(\neg DynamicLinkLibrary) \sqcap (\exists has \dash action.\{connect \mymathhyphen to \mymathhyphen ftp \mymathhyphen server\}) \sqcap (\exists has \dash action .\{enumerate \mymathhyphen registry \mymathhyphen key \mymathhyphen values\})
\end{dmath}

We can see that expressions are highly interpretable and human readable. For instance, expression \eqref{concept1} could be interpreted as follows: \textit{a binary is malicious if it has multiple sections that are executable and has at least two sections with high value of entropy}. Given that the class expressions are in fact logical formulae, they may also be used with an ontology-based retrieval system to query for samples that satisfy each given expressions.

\section{Conclusions}
\label{sec:conclusions}

The PE Ontology, described in this report, is intended as an inter-operable representation format that can be used to publish datasets of samples of PE malware. The ontology is partly based on the \acs{EMBER} dataset but does not copy the EMBER data properties one-to-one but instead relied on expert knowledge and standard nomenclature to improve interpretability and reusability.  

We hope it may also serve as a first step to establish an industry-standard for a semantic schema specifying what data should actually be included in these datasets -- considering its future gradual extensions, if needed.

It can be readily used to extract class expressions of learned malware samples either ex post or during the learning process when malware detection models are learned. As briefly hinted in Section~\ref{sec:experiments} such class expressions can than serve e.g.\ to generate explanations in human language.

The ontology, being based on EMBER, is currently suited especially for data resulting from static malware analysis. For future work, it might be useful to extend it to support data resulting from dynamic malware analysis as well.

\section{Acknowledgement}

This research was sponsored by Slovak Republic under grants APVV-19-0220 (ORBIS) and by the EU H2020 programme under Contract no.~952215 (TAILOR)

\bibliographystyle{plainnat}
\bibliography{references}  %%% Uncomment this line and comment out the ``thebibliography'' section below to use the external .bib file (using bibtex) .





% \end{thebibliography}

\end{document}